\documentclass[journal,article,submit,moreauthors]{mdpi} 
\UseRawInputEncoding
\firstpage{1} 
\hreflink{https://doi.org/} 

\usepackage{wrapfig}
\usepackage[pscoord]{eso-pic}
\usepackage[fulladjust]{marginnote}

\reversemarginpar

\definecolor{Gray}{gray}{.25}

\Title{\boldmath Scalar field in Reissner-Nordstr\"{o}m spacetime: Bound state and scattering
state } 

\Author{Shi-Lin Li $^{1}$, Yuan-Yuan Liu $^{1,3}$ , Wen-Du Li $^{2,3,\ddagger}$ , and Wu-Sheng Dai $^{1,\ddagger}$}

\address{%
$^{1}$ \quad Department of Physics, Tianjin University, Tianjin 300350, P.R. China\\
$^{2}$ \quad College of Physics and Materials Science, Tianjin Normal University, Tianjin 300387, PR China\\
$^{3}$ \quad Theoretical Physics Division, Chern Institute of Mathematics, Nankai University, Tianjin, 300071, P. R. China
}

\firstnote{liwendu@tjnu.edu.cn.} 
\secondnote{daiwusheng@tju.edu.cn.}

\abstract{In this paper, we solve the massive scalar field in the Reissner-Nordstr\"{o}m
spacetime. The scalar field in the Reissner-Nordstr\"{o}m spacetime has both
bound states and scattering states. For bound states, we solve the bound-state
wave function and the eigenvalue spectrum. For scattering states, we solve the
scattering wave function and give an explicit expression for scattering phase
shift by the integral equation method. Especially, we introduce the tortoise
coordinate for the Reissner-Nordstr\"{o}m spacetime.}


\usepackage{titletoc}   
\titlecontents{section}[0.6em]{\vspace{.25\baselineskip}}
             {\thecontentslabel\quad}{}
             {\hspace{.5em}\titlerule*[10pt]{$\cdot$}\contentspage}
\titlecontents{subsection}[2.1em]{\vspace{.25\baselineskip}}
             {\thecontentslabel\quad}{}
             {\hspace{.5em}\titlerule*[10pt]{$\cdot$}\contentspage}

\begin{document}
\nolinenumbers

\tableofcontents



\section{Introduction}

Scattering on black hole backgrounds provides insights in the evolution of
perturbations \textquotedblleft at infinity\textquotedblright%
\ \cite{angelopoulos2020non}. Scattering by black holes can help us to
understand the nature of black holes \cite{anderson2002scattering}. Scattering
in the Reissner-Nordstr\"{o}m spacetime, including scalar field scattering
\cite{benone2014absorption,crispino2009scattering}, spinor field scattering
\cite{cotuaescu2016partial,sporea2017scattering,thierry2010time}, and
electromagnetic field scattering \cite{crispino2009electromagnetic} is an
important issue. Some methods are developed for the calculation of scattering
on black holes, such as the Born approximation \cite{batic2012born}, the
integral equation method \cite{li2018scalar}, and the partial wave method
\cite{crispino2009scattering}. In scattering problems, exact solutions play
important roles, e.g., the scalar scattering in the Schwarzschild spacetime
\cite{vieira2016confluent} and in the Kerr-Newman spacetime
\cite{vieira2014exact}, the solution of the Regge-Wheeler equation, and the
solution of the Teukolsky radial equation \cite{fiziev2011application}.
Moreover, various kinds of fields, e.g., scalar fields
\cite{macedo2014absorption,batic2012orbiting,brito2013massive}, spinor fields
\cite{ahn2008black}, and vector fields
\cite{rosa2012massive,batic2012orbiting} scattered on various kinds of
spacetime, e.g., the Kerr spacetime \cite{brito2013massive}, the dyonic black
hole \cite{vieira2016confluent}, the deformed non-rotating spacetime
\cite{pei2015scattering}, and the arbitrary dimensional black holes
\cite{okawa2011super} are studied. The scattering method is important in the
calculation of the Hawking radiation, such as the Hawking radiation of the
Reissner-Nordstr\"{o}m-de Sitter black hole \cite{zhao2010hawking} and of the
Kerr-Newman black hole \cite{zhou2008hawking}, the spinor particle Hawking
radiation \cite{li2008hawking,li2008dirac}, the Hawking radiation of the
acoustic black hole \cite{zhang2011hawking}, and the species problem in the
Hawking radiation \cite{chen2018entropy}. In this paper, we determine the
scattering boundary condition by the asymptotic property of the confluent Heun
function. The Heun function is used in many physical problems, especially in
exactly solving dynamical equations
\cite{hortaccsu2013heun,ishkhanyan2016schrodinger,batic2015potentials,li2016exact,batic2013potentials,batic2018semicommuting,li2019scattering}%
.

A scalar field in the Reissner-Nordstr\"{o}m spacetime is described by the
scalar equation%
\begin{equation}
\left(  \frac{1}{\sqrt{-g}}\frac{\partial}{\partial x^{\mu}}\sqrt{-g}g^{\mu
\nu}\frac{\partial}{\partial x^{\nu}}-\mu^{2}\right)  \Phi=0 \label{KG}%
\end{equation}
under the Reissner-Nordstr\"{o}m metric:
\begin{equation}
ds^{2}=-\left(  1-\frac{2M}{r}+\frac{Q^{2}}{r^{2}}\right)  dt^{2}+\frac
{1}{\displaystyle1-\frac{2M}{r}+\frac{Q^{2}}{r^{2}}}dr^{2}+r^{2}d\Omega^{2}.
\label{metric}%
\end{equation}
The Reissner-Nordstr\"{o}m spacetime is a charged spacetime with the charge
$Q$ and the mass $M$.

A scalar field in the Reissner-Nordstr\"{o}m spacetime has both bound states
and scattering states. We first solve the bound-state solution and the
scattering-state solution by solving the scalar field equation in the
Reissner-Nordstr\"{o}m spacetime directly. By constructing the tortoise
coordinate of the Reissner-Nordstr\"{o}m spacetime, we convert the scalar
radial equation into a one-dimensional stationary Schr\"{o}dinger equation. We
solve the bound-state wave function and the eigenvalue and the scattering wave function.

Especially, we introduce the tortoise coordinate for the
Reissner-Nordstr\"{o}m spacetime for analyzing the asymptotic behavior of the field.

Moreover, in order to seek an explicit expression of the scattering phase
shift, we construct the Green function of the radial equation. Using the Green
function, we construct an integral equation for the scattering wave function.
Solving the integral equation with the scattering boundary condition, we
calculate the scattering phase shift explicitly.

Scattering in the Reissner-Nordstr\"{o}m spacetime is long-range scattering.
We need to first determine the scattering boundary condition. Determining the
scattering boundary condition for long-range scattering is often difficult,
since different long-range scatterings have different boundary conditions
\cite{hod2013scattering,liu2014scattering,li2016scattering,li2016exact}. As a
contrast, the scattering boundary condition for short-range scatterings is the
same. In this paper, we first rewrite the radial equation as a confluent Heun
equation, and then we determine the scattering boundary condition based on the
asymptotic behavior of the confluent Heun function.

Furthermore, when calculating scattering cross sections, we encounter a
difficulty in the partial sum approximation. An incorrect oscillation appears
when truncating the partial wave expansion. We show how to eliminate such an
oscillation in \cite{li2021eliminating}. 

In section \ref{fieldeq}, we solve bound-state and scattering-state solutions
directly from the scalar field equation in the Reissner-Nordstr\"{o}m
spacetime. In section \ref{Integraleq}, we solve the scattering phase shift
explicitly by the integral equation method. The conclusions are summarized in
section \ref{Conclusions}.

\section{Scalar field in Reissner-Nordstr\"{o}m spacetime: bound state and
scattering state \label{fieldeq}}

\subsection{Radial equation \label{radial}}

The scalar equation (\ref{KG}) with the Reissner-Nordstr\"{o}m metric
(\ref{metric}) leads the radial equation:%
\end{paracol}
\begin{align}
 \qquad \qquad \qquad \qquad\left\{  \left(  1-\frac{2M}{r}+\frac{Q^{2}}{r^{2}}\right)  \frac{1}{r^{2}%
}\frac{d}{dr}r^{2}\left(  1-\frac{2M}{r}+\frac{Q^{2}}{r^{2}}\right)  \frac
{d}{dr}\right. 
 +\left.  \omega^{2}-\left(  1-\frac{2M}{r}+\frac{Q^{2}}{r^{2}}\right)
\left[  \frac{l\left(  l+1\right)  }{r^{2}}+\mu^{2}\right]  \right\}
R_{l}\left(  r\right)  =0, \label{radialeq}%
\end{align}
\begin{paracol}{2}
\switchcolumn
\noindent where $R_{l}\left(  r\right)  $ is the radial wave function, $\omega^{2}$ is
energy of the incident particle, and $\mu$\ is the mass of the particle.

The Reissner-Nordstr\"{o}m spacetime has two horizons at%

\begin{equation}
r_{\pm}=M\pm\sqrt{M^{2}-Q^{2}}. \qquad\qquad\qquad\qquad\qquad\qquad\qquad\qquad\quad \label{rm}%
\end{equation}
The radial equation (\ref{radialeq}) can be reexpressed by $r_{+}$ and $r_{-}$
as%
\end{paracol}
\begin{align}
\left\{  \left(  1-\frac{r_{+}}{r}\right)  \left(  1-\frac{r_{-}}%
{r}\right)  \frac{1}{r^{2}}\frac{d}{dr}r^{2}\left(  1-\frac{r_{+}}{r}\right)
\left(  1-\frac{r_{-}}{r}\right)  \frac{d}{dr}\right. 
+\left.  \left[  \eta^{2}-\left(  1-\frac{r_{+}}{r}\right)  \left(
1-\frac{r_{-}}{r}\right)  \frac{l\left(  l+1\right)  }{r^{2}}-\left(
-\frac{r_{+}}{r}-\frac{r_{-}}{r}+\frac{r_{+}r_{-}}{r^{2}}\right)  \mu
^{2}\right]  \right\}  R_{l}\left(  r\right)  =0, \label{radialeq4}%
\end{align}
\begin{paracol}{2}
\switchcolumn
\noindent where $\eta^{2}=\omega^{2}-\mu^{2}$.

\subsection{Confluent Heun equation \label{CHeunEq}}

The radial equation (\ref{radialeq4}), by the replacement%
\begin{equation}
z=\frac{2r-\left(  r_{+}+r_{-}\right)  }{r_{+}-r_{-}} \qquad\qquad\qquad\qquad\qquad\qquad\qquad\qquad\qquad \label{zr}%
\end{equation}
and%
\begin{align}
p  &  =-\frac{i}{2}\eta\left(  r_{+}-r_{-}\right)  ,\nonumber\\
\beta &  =i\frac{r_{+}+r_{-}}{2\eta}\left(  2\eta^{2}+\mu^{2}\right)
,\nonumber\\
m  &  =\frac{r_{+}^{2}+r_{-}^{2}}{r_{+}-r_{-}}\sqrt{-\eta^{2}-\mu^{2}%
},\nonumber\\
s  &  =\left(  r_{+}+r_{-}\right)  \sqrt{-\eta^{2}-\mu^{2}},\nonumber\\
\lambda &  =l\left(  l+1\right)  -2\left(  r_{+}^{2}+r_{+}r_{-}+r_{-}%
^{2}\right)  \eta^{2}-\frac{1}{2}\left(  3r_{+}^{2}+4r_{+}r_{-}+3r_{-}%
^{2}\right)  \mu^{2}%
\end{align}
can be rewritten as%
\begin{equation}
\frac{d}{dz}\left(  z^{2}-1\right)  \frac{d}{dz}y\left(  z\right)  +\left[
-p^{2}\left(  z^{2}-1\right)  +2p\beta z-\lambda-\frac{m^{2}+s^{2}+2msz}%
{z^{2}-1}\right]  y\left(  z\right)  =0. \label{eqyz}%
\end{equation}
This is just the confluent Heun equation \cite{ronveaux1995heun}.

The radial wave function, by Eq. (\ref{zr}), can be obtained by
\begin{equation}
R_{l}\left(  r\right)  =\left.  y\left(  z\right)  \right\vert _{z=\left[
2r-\left(  r_{+}+r_{-}\right)  \right]  /\left(  r_{+}-r_{-}\right)  }.
\end{equation}

\subsection{Tortoise coordinate for Reissner-Nordstr\"{o}m spacetime}

We introduce the tortoise coordinate for the Reissner-Nordstr\"{o}m spacetime
as
\begin{align}
r_{\ast}  &  =\int\frac{dr}{\displaystyle\left(  1-\frac{r_{+}}{r}\right)
\left(  1-\frac{r_{-}}{r}\right)  }\nonumber\\
&  =r+\frac{r_{+}^{2}}{r_{+}-r_{-}}\ln\left(  \frac{r}{r_{+}}-1\right)
-\frac{r_{-}^{2}}{r_{+}-r_{-}}\ln\left(  \frac{r}{r_{-}}-1\right)  .
\label{tortoise1}%
\end{align}
Introducing $R_{l}\left(  r\right)  =u_{l}\left(  r\right)  /r$ and rewriting
the radial equation (\ref{radialeq}) under the tortoise coordinate give%
\end{paracol}
\begin{align}
 \qquad\qquad\qquad\frac{d^{2}u_{l}\left(  r\right)  }{dr_{\ast}^{2}}+\left\{  \left(
\omega^{2}-\mu^{2}\right)  -\left(  1-\frac{r_{+}}{r}\right)  \left(
1-\frac{r_{-}}{r}\right)  \left[  \frac{l\left(  l+1\right)  }{r^{2}}+\left(
\frac{r_{+}+r_{-}}{r^{3}}-\frac{2r_{+}r_{-}}{r^{4}}\right)  \right]  \right.
 -\left.  \mu^{2}\left(  \frac{r_{+}r_{-}}{r^{2}}-\frac{r_{+}+r_{-}}%
{r}\right)  \right\}  u_{l}\left(  r\right)  =0. \label{radialeq1}%
\end{align}
\begin{paracol}{2}
\switchcolumn

\subsection{Boundary condition}

For the Reissner-Nordstr\"{o}m spacetime, we need to impose three boundary
conditions at the outer and inner horizons $r=r_{\pm}$ and at $r\rightarrow
\infty$, respectively.

At the outer and inner horizons, $r=r_{\pm}$, we require that $\left\vert
R_{l}\left(  r_{\pm}\right)  \right\vert $ is finite:%
\begin{equation}
\left\vert R_{l}\left(  r=r_{\pm}\right)  \right\vert <\infty.
\label{boundaryRrprm}%
\end{equation}
The boundary condition at $r\rightarrow\infty$ needs to be considered in two
cases: scattering states and bound states. For bound states, the boundary
condition is%
\begin{equation}
\left\vert R_{l}\left(  r\rightarrow\infty\right)  \right\vert =0.
\label{boundaryinfbound}%
\end{equation}
For scattering states, the boundary condition is determined by the asymptotics
behavior and satisfies
\begin{equation}
\left\vert R_{l}\left(  r\rightarrow\infty\right)  \right\vert <\infty.
\label{boundaryinfsca}%
\end{equation}

We determine the boundary conditions at $r\rightarrow\infty$ and $r\rightarrow
r_{\pm}$ by analyzing the asymptotic behavior of the radial equation, respectively.

\textit{The boundary condition at the outer horizon.} At the outer horizon,
$r=r_{+}$, the asymptotics of the radial equation for $r\rightarrow r_{+}$ by
Eq. (\ref{radialeq1}) reads
\begin{equation}
\frac{d^{2}u_{l}\left(  r\right)  }{dr_{\ast}^{2}}+\omega^{2}u_{l}\left(
r\right)  \overset{r\rightarrow r_{+}}{\sim}0. \label{aymrplus}%
\end{equation}
The solution of the asymptotic equation (\ref{aymrplus}) is%
\begin{equation}
u_{l}\left(  r\right)  \overset{r\rightarrow r_{+}}{\sim}e^{\pm i\omega
r_{\ast}}.
\end{equation}
That is, the scattering boundary at $r\rightarrow r_{+}$ is%
\begin{equation}
R_{l}\left(  r\right)  \overset{r\rightarrow r_{+}}{\sim}\frac{1}{r_{+}}e^{\pm
i\omega r_{\ast}}\sim e^{\pm i\omega r_{\ast}}.
\end{equation}

\textit{The boundary condition at the inner horizon. }At the inner horizon,
$r=r_{-}$, the asymptotics of the radial equation for $r\rightarrow r_{-}$ by
Eq. (\ref{radialeq1}) reads%
\begin{equation}
\frac{d^{2}u_{l}\left(  r\right)  }{dr_{\ast}^{2}}+\omega^{2}u_{l}\left(
r\right)  \overset{r\rightarrow r_{-}}{\sim}0. \label{asymptoticeq}%
\end{equation}
The solution of the asymptotic equation (\ref{asymptoticeq}) is%
\begin{equation}
u_{l}\left(  r\right)  \overset{r\rightarrow r_{-}}{\sim}e^{\pm i\omega
r_{\ast}}.
\end{equation}
That is, the scattering boundary at $r\rightarrow r_{-}$ is%
\begin{equation}
R_{l}\left(  r\right)  \overset{r\rightarrow r_{-}}{\sim}\frac{1}{r_{-}}e^{\pm
i\omega r_{\ast}}\sim e^{\pm i\omega r_{\ast}}.
\end{equation}

\textit{The bound-state boundary condition. }The bound-state boundary
condition is
\begin{equation}
R_{l}\left(  r\right)  \overset{r\rightarrow\infty}{\sim}0;
\end{equation}
that is, $u_{l}\left(  r\right)  $ is finite at $r\rightarrow\infty$.

\textit{The scattering boundary condition. }The asymptotics of the radial
equation for $r\rightarrow\infty$ by Eq. (\ref{radialeq1}) reads%
\begin{equation}
\frac{d^{2}u_{l}\left(  r\right)  }{dr_{\ast}^{2}}+\eta^{2}u_{l}\left(
r\right)  \overset{r\rightarrow\infty}{\sim}0. \label{asymptoticeq1}%
\end{equation}
The solution of the asymptotic equation (\ref{asymptoticeq1}) is%
\begin{equation}
u_{l}\left(  r\right)  \overset{r\rightarrow\infty}{\sim}e^{\pm i\eta r_{\ast
}}.
\end{equation}
That is, the scattering boundary at $r\rightarrow\infty$ is
\begin{equation}
R_{l}\left(  r\right)  \overset{r\rightarrow\infty}{\sim}\frac{1}{r}e^{\pm
i\eta r_{\ast}}.
\end{equation}

In section \ref{CHeunEq}, we convert the radial equation (\ref{radialeq}) into
the confluent Heun equation (\ref{eqyz}), so the boundary conditions of
$R_{l}\left(  r\right)  $ should be converted into the boundary conditions of
$y(z)$.

The boundary conditions at the outer and inner horizons, Eq.
(\ref{boundaryRrprm}), becomes%
\begin{equation}
\left\vert y\left(  z=\pm1\right)  \right\vert <\infty.
\end{equation}
The boundary conditions at $r\rightarrow\infty$, Eqs. (\ref{boundaryinfsca})
and (\ref{boundaryinfbound}) become%
\begin{align}
\left\vert y\left(  z\rightarrow\infty\right)  \right\vert  &  <\infty,\text{
\ scattering state,}\nonumber\\
\left\vert y\left(  z\rightarrow\infty\right)  \right\vert  &  =0,\text{
\ bound state.}%
\end{align}

\subsection{Bound-state solution}

In section \ref{CHeunEq}, we convert the radial equation (\ref{radialeq4})
into the confluent Heun equation (\ref{eqyz}).

For bound states%
\begin{equation}
\eta=ik,\text{ \ } \quad k>0 \qquad\qquad\qquad\qquad\qquad\qquad\qquad\qquad\quad\quad\quad \label{etak}%
\end{equation}
and
\begin{align}
p  &  =\frac{1}{2}k\left(  r_{+}-r_{-}\right)  ,\\
\beta &  =\frac{r_{+}+r_{-}}{2k}\left(  -2k^{2}+\mu^{2}\right)  ,\\
m  &  =\frac{r_{+}^{2}+r_{-}^{2}}{r_{+}-r_{-}}\sqrt{k^{2}-\mu^{2}},\\
s  &  =\left(  r_{+}+r_{-}\right)  \sqrt{k^{2}-\mu^{2}},\\
\lambda &  =l\left(  l+1\right)  +2\left(  r_{+}^{2}+r_{+}r_{-}+r_{-}%
^{2}\right)  k^{2}-\frac{1}{2}\left(  3r_{+}^{2}+4r_{+}r_{-}+3r_{-}%
^{2}\right)  \mu^{2}. \label{lamdak}%
\end{align}
Here $\eta=-k^{2}$ in Eq. (\ref{etak}) is the eigenvalue of the radial
equation (\ref{radialeq4}). The bound-state boundary condition for the radial
equation $R\left(  r\right)  \overset{r\rightarrow\infty}{\sim}0$ corresponds
to the boundary condition for the confluent Heun equation (\ref{eqyz})
$y\left(  z\right)  \overset{z\rightarrow\infty}{\sim}0$. The boundary
conditions at the horizons $r_{\pm}$ correspond to the boundary conditions of
the confluent Heun equation (\ref{eqyz}) at $z=\pm1$. That is, we need a
solution of the confluent Heun equation (\ref{eqyz}) satisfying $\left\vert
\Pi\left(  p,\beta,\pm1\right)  \right\vert <\infty$ and $\Pi\left(
p,\beta,z\right)  \overset{z\rightarrow\infty}{\sim}0$. Such a solution of Eq.
(\ref{eqyz}) is \cite{ronveaux1995heun}
\end{paracol}
\begin{align}
\qquad\qquad\qquad\qquad\Pi\left(  p,\beta,z\right)      =N\left(  z-1\right)  ^{\left(  m+s\right)
/2}\left(  z+1\right)  ^{\left(  m-s\right)  /2}e^{-p\left(  1+z\right)
} \operatorname*{Hc}{}^{\left(  a\right)  }\left(  p,-\beta
+m+1,m+s+1,m-s+1,\sigma;\frac{z+1}{2}\right)  ,
\end{align}
\begin{paracol}{2}
\switchcolumn
\noindent where $\sigma=\lambda+2p\left(  -2\beta+m+s+1\right)  -m\left(  m+1\right)  $
and $\operatorname*{Hc}{}^{\left(  a\right)  }\left(  a,b,c,d,e;z\right)  $ is
the confluent Heun function. Then the solution of the radial equation reads%
\end{paracol}
\begin{align}
u_{l}\left(  r\right)   &  =4Ne^{-k\left(  r-r_{-}\right)  }\left(
\frac{r-r_{+}}{r_{+}-r_{-}}\right)  ^{\frac{r_{+}^{2}+r_{-}^{2}}{r_{+}-r_{-}%
}\sqrt{k^{2}-\mu^{2}}}\nonumber\\
&  \times\operatorname*{Hc}{}^{\left(  a\right)  }\left(  \frac{1}{2}k\left(
r_{+}-r_{-}\right)  ,\left(  r_{+}+r_{-}\right)  \frac{2k^{2}-\mu^{2}}%
{2k}+\frac{r_{+}^{2}+r_{-}^{2}}{r_{+}-r_{-}}\sqrt{k^{2}-\mu^{2}}+1,\right.
  \left.  \frac{2r_{+}^{2}}{r_{+}-r_{-}}\sqrt{k^{2}-\mu^{2}}+1,\frac
{2r_{-}^{2}}{r_{+}-r_{-}}\sqrt{k^{2}-\mu^{2}}+1,\sigma;\frac{2r-\left(
r_{+}+r_{-}\right)  +1}{2\left(  r_{+}-r_{-}\right)  }\right)  ,
\label{boundsW}%
\end{align}
\begin{paracol}{2}
\switchcolumn
\noindent
where%
\end{paracol}
\begin{align}
\qquad\qquad\qquad\qquad\sigma   = & l\left(  l+1\right)  +2\left(  r_{+}^{2}+r_{+}r_{-}+r_{-}%
^{2}\right)  k^{2}-\frac{1}{2}\left(  3r_{+}^{2}+4r_{+}r_{-}+3r_{-}%
^{2}\right)  \mu^{2}
  +k\left(  r_{+}-r_{-}\right)  \left[  \left(  r_{+}+r_{-}\right)
\frac{2k^{2}-\mu^{2}}{k}+\frac{2r_{+}^{2}}{r_{+}-r_{-}}\sqrt{k^{2}-\mu^{2}%
}+1\right]  \nonumber\\ 
& -\frac{r_{+}^{2}+r_{-}^{2}}{r_{+}-r_{-}}\sqrt{k^{2}-\mu^{2}}\left(
\frac{r_{+}^{2}+r_{-}^{2}}{r_{+}-r_{-}}\sqrt{k^{2}-\mu^{2}}+1\right)  .
\end{align}

\begin{paracol}{2}
\switchcolumn
The eigenvalue of the radial equation (\ref{radialeq4}), $k^{2}$, relates the
eigenvalue of the confluent Heun equation (\ref{eqyz}), $\lambda_{n}$, through
the relation (\ref{lamdak}). By the relation (\ref{lamdak}), we arrive at%
\begin{equation}
k^{2}=\frac{\lambda_{n}+\frac{1}{2}\left(  3r_{+}^{2}+4r_{+}r_{-}+3r_{-}%
^{2}\right)  \mu^{2}-l\left(  l+1\right)  }{2\left(  r_{+}^{2}+r_{+}%
r_{-}+r_{-}^{2}\right)  }. \qquad\qquad\qquad\qquad\qquad\quad \label{kn}%
\end{equation}

For a large $p=\frac{1}{2}k\left(  r_{+}-r_{-}\right)  $, the eigenvalue of
the confluent Heun equation (\ref{eqyz}), $\lambda_{n}$, has the following
asymptotics \cite{ronveaux1995heun},%
\begin{align}
\lambda_{n}  &  =k\left(  r_{+}-r_{-}\right)  \left[  2\chi_{n}+\left(
r_{+}+r_{-}\right)  \frac{2k^{2}-\mu^{2}}{2k}\right]  -2\chi\left[
\chi+\left(  r_{+}+r_{-}\right)  \frac{2k^{2}-\mu^{2}}{2k}\right] \nonumber\\
&  +\frac{1}{2}\left\{  \left[  \left(  \frac{r_{+}^{2}+r_{-}^{2}}{r_{+}%
-r_{-}}\right)  ^{2}+\left(  r_{+}+r_{-}\right)  ^{2}\right]  \left(
k^{2}-\mu^{2}\right)  -1\right\}  +O\left(  \frac{1}{k\left(  r_{+}%
-r_{-}\right)  /2}\right)  , \label{lambdap}%
\end{align}
where
\begin{equation}
\chi_{n}=n+\frac{1}{2} \label{kain}%
\end{equation}
with $n$ an integer.

Then the eigenvalue of the radial equation $k^{2}$ can be solved from Eqs.
(\ref{kn}) and (\ref{lambdap}):
\begin{equation}
k=\frac{1}{3A^{\prime}}\left[  \left(  \frac{\sqrt{E^{\prime2}+4F^{\prime3}%
}+E^{\prime}}{2}\right)  ^{1/3}-F^{\prime}\left(  \frac{\sqrt{E^{\prime
2}+4F^{\prime3}}+E^{\prime}}{2}\right)  ^{-1/3}-B^{\prime}\right]  ,
\end{equation}
where%
\begin{align}
A^{\prime}  &  =-\frac{\left(  r_{+}^{2}+2r_{+}r_{-}-r_{-}^{2}\right)  \left(
r_{+}^{2}-2r_{+}r_{-}+3r_{-}^{2}\right)  }{2\left(  r_{+}-r_{-}\right)  ^{2}%
}\nonumber\\
&  =\frac{4M^{4}-8M^{2}Q^{2}+3Q^{4}-8M\left(  M^{2}-Q^{2}\right)  ^{3/2}%
}{2\left(  M^{2}-Q^{2}\right)  },\qquad\qquad\qquad\qquad\qquad\qquad\qquad \nonumber\\
B^{\prime}  &  =2\left(  2n+1\right)  r_{-}=2\left(  2n+1\right)  \left(
M-\sqrt{M^{2}-Q^{2}}\right) , \qquad\qquad\qquad\qquad\qquad\qquad\qquad
\end{align}%
\begin{align}
E^{\prime}  &  =\frac{2n+1}{8}\left\{  -\frac{36\left(  r_{+}^{2}+2r_{+}%
r_{-}-r_{-}^{2}\right)  \left(  r_{+}^{2}-2r_{+}r_{-}+3r_{-}^{2}\right)
r_{-}}{\left(  r_{+}-r_{-}\right)  ^{4}}\right. \nonumber\\
&  \times\left\{  \left[  2l\left(  l+1\right)  +4n\left(  n+1\right)
+3\right]  (r_{+}-r_{-})^{2}+\left(  r_{+}^{4}+4r_{+}^{3}r_{-}-r_{-}%
^{4}\right)  \mu^{2}\right\} \nonumber\\
&  \left.  -128\left(  2n+1\right)  ^{2}r_{-}^{3}+\frac{27\left(  r_{+}%
+r_{-}\right)  \left(  r_{+}^{2}+2r_{+}r_{-}-r_{-}^{2}\right)  ^{2}\left(
r_{+}^{2}-2r_{+}r_{-}+3r_{-}^{2}\right)  ^{2}}{\left(  r_{+}-r_{-}\right)
^{4}}\mu^{2}\right\} \nonumber\\
&  =\frac{2n+1}{8}\left\{  -128(2n+1)^{2}\left(  M-\sqrt{M^{2}-Q^{2}}\right)
^{3}\right. \nonumber\\
&  \left.  -\frac{36\left[  -4M^{4}+8M^{2}Q^{2}-3Q^{4}+8M\left(  M^{2}%
-Q^{2}\right)  ^{3/2}\right]  \left(  M-\sqrt{M^{2}-Q^{2}}\right)  }{\left(
M^{2}-Q^{2}\right)  ^{2}}\right. \nonumber\\
&  \left.  \times\left\{  \left[  4n\left(  n+1\right)  +2l\left(  l+1\right)
+3\right]  \left(  M^{2}-Q^{2}\right)  +\left(  2M^{2}-Q^{2}\right)  Q^{2}%
\mu^{2}+4M\left(  M^{2}-Q^{2}\right)  ^{3/2}\mu^{2}\right\}  \right.
\nonumber\\
&  \left.  +\frac{54M\left[  4M^{4}-8M^{2}Q^{2}+3Q^{4}-8M\left(  M^{2}%
-Q^{2}\right)  ^{3/2}\right]  ^{2}}{\left(  M^{2}-Q^{2}\right)  ^{2}}\mu
^{2}\right\}  ,
\end{align}%
\begin{align}
F^{\prime}  &  =-\frac{3\left(  r_{+}^{2}+2r_{+}r_{-}-r_{-}^{2}\right)
\left(  r_{+}^{2}-2r_{+}r_{-}+3r_{-}^{2}\right)  }{4\left(  r_{+}%
-r_{-}\right)  ^{4}}\nonumber\\
&  \times\left\{  \left[  2l\left(  l+1\right)  +4n\left(  n+1\right)
+3\right]  (r_{+}-r_{-})^{2}+\left(  r_{+}^{4}+4r_{+}^{3}r_{-}-r_{-}%
^{4}\right)  \mu^{2}\right\}  -4\left(  2n+1\right)  ^{2}r_{-}^{2}\nonumber\\
&  =\frac{3}{4\left(  M^{2}-Q^{2}\right)  ^{2}}\left[  4M^{4}-8M^{2}%
Q^{2}+3Q^{4}-8M\left(  M^{2}-Q^{2}\right)  ^{3/2}\right] \nonumber\\
&  \times\left\{  \left[  4n\left(  n+1\right)  +2l\left(  l+1\right)
+3\right]  \left(  M^{2}-Q^{2}\right)  +\left(  2M^{2}-Q^{2}\right)  Q^{2}%
\mu^{2}+4M\left(  M^{2}-Q^{2}\right)  ^{3/2}\mu^{2}\right\} \nonumber\\
&  -4(2n+1)^{2}\left(  M-\sqrt{M^{2}-Q^{2}}\right)  ^{2}.
\end{align}
\ \ \ \ \ \ \ \ \ 

The eigenvalue of the radial equation then reads
\begin{equation}
\omega^{2}=\mu^{2}-k^{2}.
\end{equation}

For large $k$, the expansion of Eqs. (\ref{lamdak}) and (\ref{lambdap}) gives%
\begin{equation}
k=-\frac{2n\left(  n+1\right)  +l\left(  l+1\right)  +1-\sqrt{\left[
2n\left(  n+1\right)  +l\left(  l+1\right)  +1\right]  ^{2}+4\left(
2n+1\right)  ^{2}Q^{2}\mu^{2}}}{2\left(  2n+1\right)  }\frac{M}{Q^{2}}%
\end{equation}
and the eigenvalue%
\end{paracol}
\begin{align}
\qquad\qquad\qquad\qquad\qquad\omega  =\sqrt{\mu^{2}-k^{2}} 
=\sqrt{\mu^{2}-\left\{  \frac{2n\left(  n+1\right)  +l\left(  l+1\right)
+1-\sqrt{\left[  2n\left(  n+1\right)  +l\left(  l+1\right)  +1\right]
^{2}+4\left(  2n+1\right)  ^{2}Q^{2}\mu^{2}}}{2\left(  2n+1\right)  }\frac
{M}{Q^{2}}\right\}  ^{2}}. \label{omg}%
\end{align}

\begin{paracol}{2}
\switchcolumn
For a large $n$, i.e., the high-energy case, the eigenvalue becomes%
\begin{equation}
\omega^{2}=\mu^{2}-k^{2}=\mu^{2}-\frac{\mu^{4}M^{2}}{n^{2}}+\frac{2\mu
^{6}M^{2}Q^{2}}{n^{4}}. \label{omega3}%
\end{equation}
That is, the contribution of the charge $Q$ is proportional to $n^{-4}$, while
the contribution of the mass $M$ is proportional to $n^{-2}$. For a large $n$
or a small $Q$, the eigenvalue (\ref{omega3}) becomes%
\begin{equation}
\omega^{2}=\mu^{2}-k^{2}=\mu^{2}-\frac{\mu^{4}M^{2}}{n^{2}}.
\end{equation}
This agrees with the result of a Klein-Gordon particle in the Coulomb
potential in the high-energy case. The eigenvalue of the Klein-Gordon equation
with the Coulomb potential is $\eta_{\text{Coulomb}}^{2}=\mu^{2}\left[
1+\alpha^{2}/\left(  n+\beta\right)  ^{2}\right]  ^{-1/2}-\mu^{2}$, where
$\alpha$ and $\beta$ are some constants \cite{schwabl2013advanced}. For a
large $n$, $\eta_{\text{Coulomb}}^{2}\sim-1/n^{2}$ \cite{li2019scattering}.

For small $n$, i.e., the low-energy case, for a large $Q$, consequently a
large $M$ for $Q$ must be less than $M$, the eigenvalue becomes%
\begin{equation}
\omega^{2}=\mu^{2}-k^{2}=\left\{  \mu^{2}-\frac{M\left[  l\left(  l+1\right)
+1-2Q\mu\right]  }{2Q^{2}}\right\}  -\frac{Ml\left(  l+1\right)  }{Q^{2}}n.
\end{equation}
This agrees with the result of a Klein-Gordon particle in the Coulomb
potential in the low-energy case. For small $n$, the eigenvalue of the Coulomb
potential is $\eta_{\text{Coulomb}}^{2}=A-Bn$ \cite{li2019scattering}.

When the charge $Q\rightarrow0$, the eigenvalue\ (\ref{omg}) becomes
\begin{equation}
\omega^{2}=\mu^{2}-k^{2}=\mu^{2}-\left[  \frac{2n+1}{2n\left(  n+1\right)
+l\left(  l+1\right)  +1}\right]  ^{2}\mu^{4}M^{2}, \label{omega2}%
\end{equation}
which agrees with the corresponding result in the Schwarzschild spacetime
given by Ref. \cite{li2019scattering}.

For illustration, we rearrange the expression of the eigenvalue, Eq.
(\ref{omg}), as%
\begin{equation}
\frac{\omega}{\mu}=\sqrt{1-\left\{  \frac{1-\sqrt{1+\left[  \frac{2\left(
2n+1\right)  }{2n\left(  n+1\right)  +l\left(  l+1\right)  +1}\left(  \frac
{Q}{M}\right)  \left(  M\mu\right)  \right]  ^{2}}}{2\left(  2n+1\right)
/\left[  2n\left(  n+1\right)  +l\left(  l+1\right)  +1\right]  }\frac{1}%
{M\mu\left(  \frac{Q}{M}\right)  ^{2}}\right\}  ^{2}}.
\end{equation}
In Figs. (\ref{figeigen}), (\ref{Mmu}), and (\ref{QM}) we plot the eigenvalue
as a function of the mass and the charge.

\begin{wrapfigure}[18]{l}[0.5cm]{0.5\textwidth}
\includegraphics[width=0.5\textwidth]{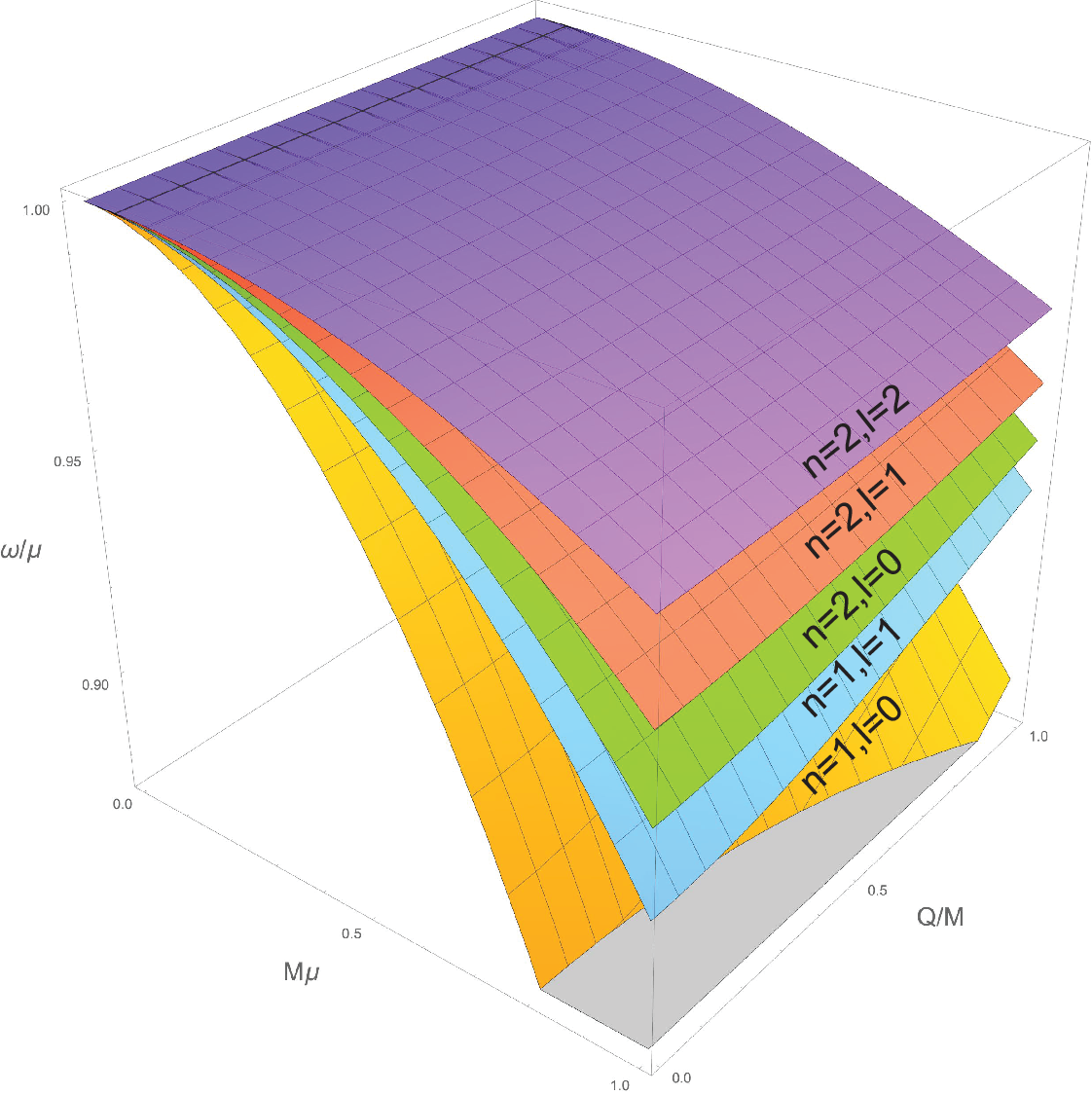}
\captionsetup{labelformat=empty} 
\caption{} 
\label{figeigen} 
\end{wrapfigure}
\
\qquad\qquad\\
\qquad\qquad\\
\qquad\qquad\\
\qquad\qquad\\
\qquad\qquad\\
\qquad\qquad\\
\qquad\qquad\\
\qquad\qquad\\
\qquad\qquad\\
\vspace*{1.5cm}
\qquad\qquad\\
\qquad\qquad\\
\qquad\qquad\\
\qquad\qquad\\
\qquad\qquad\\
\qquad\qquad\\
\qquad\qquad\\

\bigskip
\switchcolumn
\newpage
\vspace*{8cm}
{\color{Gray} 
\noindent\textbf{Figure\ref{figeigen}.} The bound-state eigenvalue
$\omega/\mu$ as a function of $M\mu$ and $Q/M$. 
}
\switchcolumn

\vspace*{0.5cm}
\begin{wrapfigure}[18]{l}[0.5cm]{0.5\textwidth}
\includegraphics[width=0.5\textwidth]{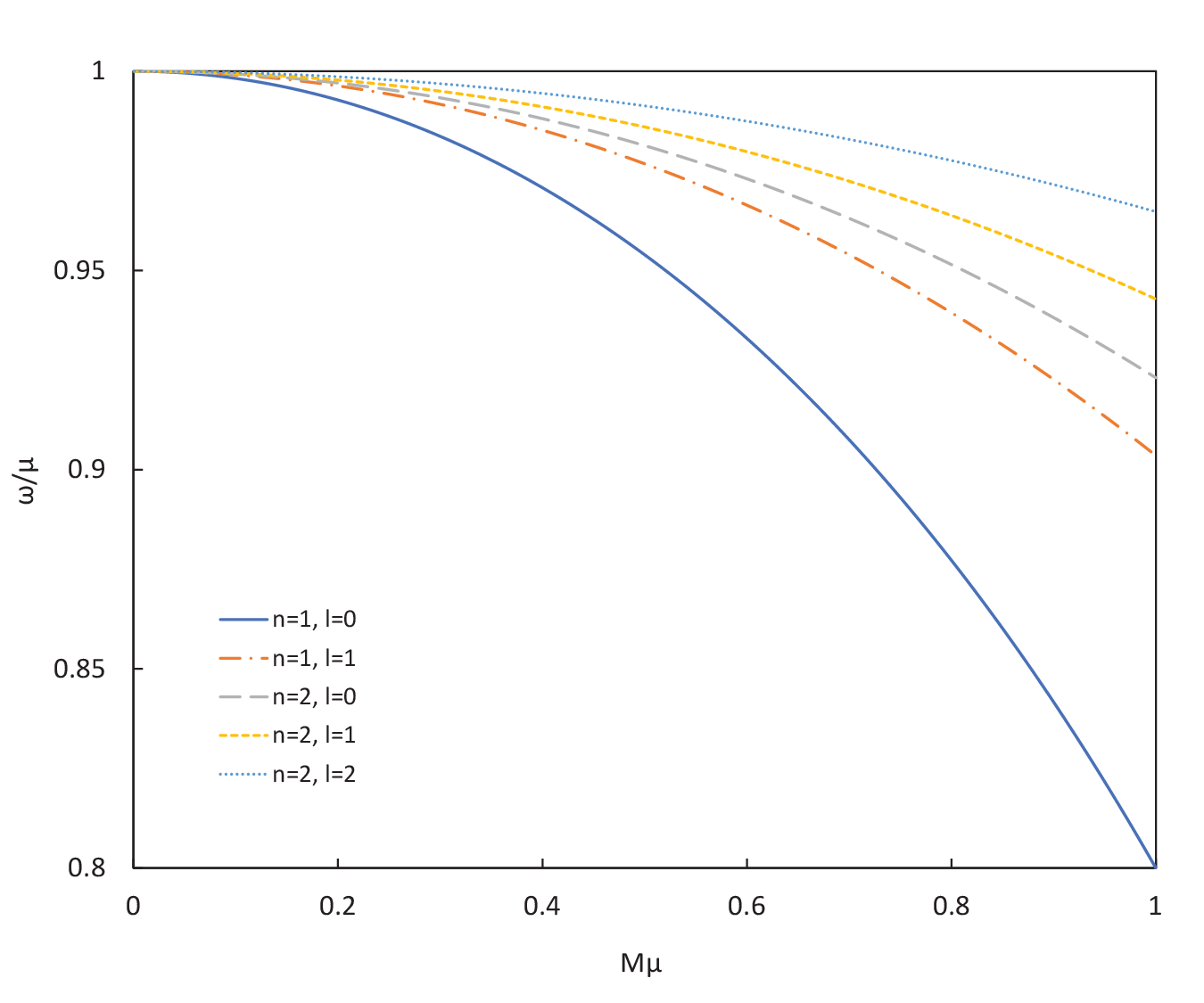}
\captionsetup{labelformat=empty} 
\caption{} 
\label{Mmu} 
\end{wrapfigure}
\
\qquad\qquad\\
\qquad\qquad\\
\qquad\qquad\\
\qquad\qquad\\
\qquad\qquad\\
\qquad\qquad\\
\qquad\qquad\\
\qquad\qquad\\
\qquad\qquad\\
\vspace*{0.5cm}
\qquad\qquad\\
\qquad\qquad\\
\qquad\qquad\\
\qquad\qquad\\
\qquad\qquad\\
\qquad\qquad\\
\qquad\qquad\\

\bigskip
\switchcolumn
\vspace*{8.5cm}
{\color{Gray} 
\noindent\textbf{Figure \ref{Mmu}.} The bound-state
eigenvalue $\omega/\mu$ as a function of $M\mu$ with $Q/M=10^{-3}$. 
}
\switchcolumn

\vspace*{-3.0cm}
\begin{wrapfigure}[18]{l}[0.5cm]{0.5\textwidth}
\includegraphics[width=0.5\textwidth]{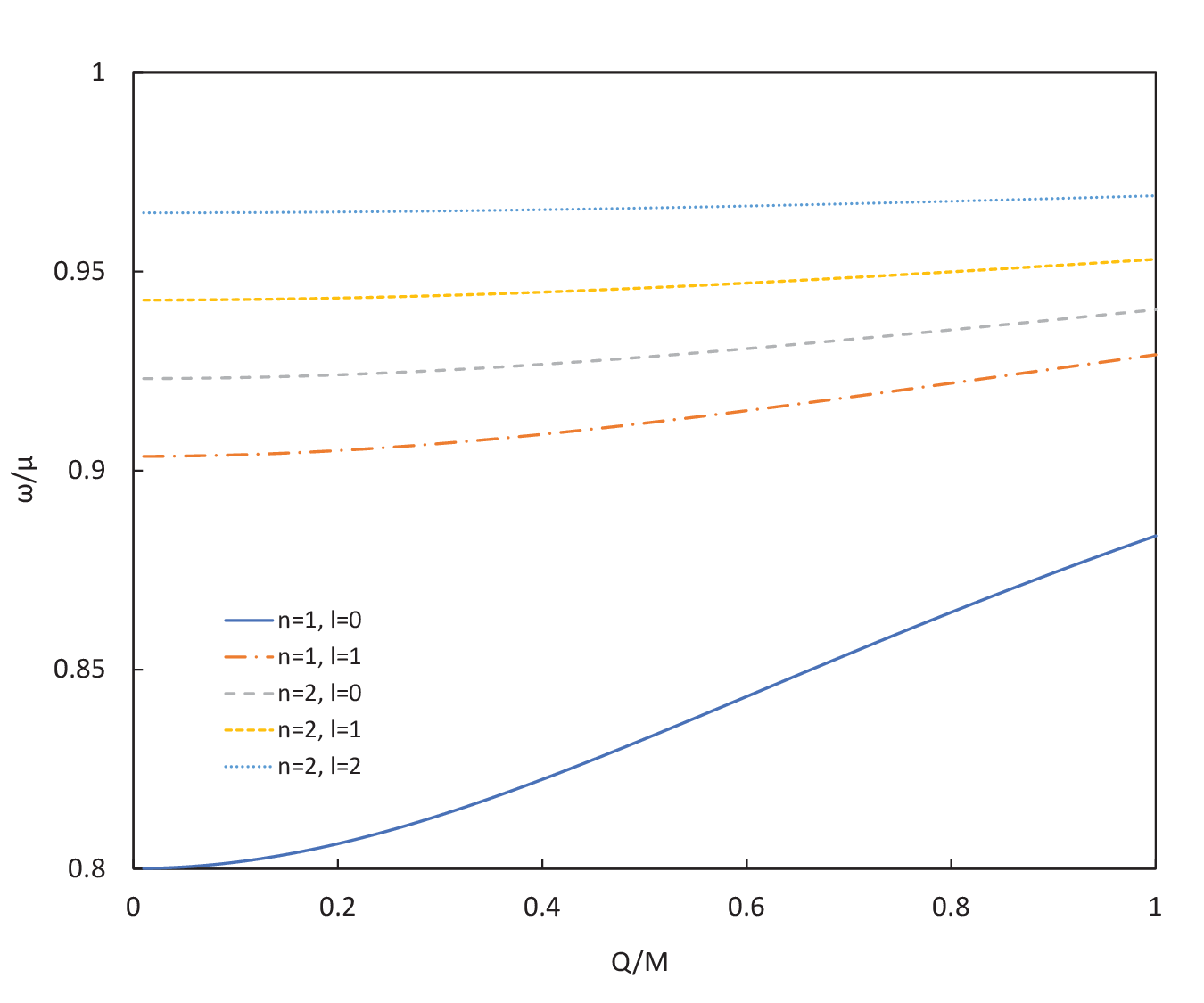}
\captionsetup{labelformat=empty} 
\caption{} 
\label{QM} 
\end{wrapfigure}
\
\qquad\qquad\\
\qquad\qquad\\
\qquad\qquad\\
\qquad\qquad\\
\qquad\qquad\\
\qquad\qquad\\
\qquad\qquad\\
\qquad\qquad\\
\qquad\qquad\\
\vspace*{2.0cm}
\qquad\qquad\\
\qquad\qquad\\
\qquad\qquad\\
\qquad\qquad\\
\qquad\qquad\\
\qquad\qquad\\
\qquad\qquad\\

\bigskip
\switchcolumn
\newpage
\vspace*{1.0cm}
{\color{Gray} 
\noindent\textbf{Figure \ref{QM}.} The bound-state eigenvalue
$\omega/\mu$ as a function of $Q/M$ with $M\mu=1$. 
}
\switchcolumn
\vspace*{-1.0cm}
\subsection{Scattering solution}

The scattering solution can be obtained by performing the replacement
\begin{equation}
\eta=ik
\end{equation}
in Eq. (\ref{boundsW}):%
\end{paracol}
\begin{align}
u_{l}\left(  r\right)   &  =4Ne^{i\eta\left(  r-r_{-}\right)  }\left(
\frac{r-r_{+}}{r_{+}-r_{-}}\right)  ^{i\frac{r_{+}^{2}+r_{-}^{2}}{r_{+}-r_{-}%
}\sqrt{\eta^{2}+\mu^{2}}}\nonumber\\
&  \times\operatorname*{Hc}{}^{\left(  a\right)  }\left(  -\frac{i}{2}%
\eta\left(  r_{+}-r_{-}\right)  ,-\frac{i\left(  r_{+}+r_{-}\right)  \left(
2\eta^{2}+\mu^{2}\right)  }{2\eta}+\frac{i\left(  r_{+}^{2}+r_{-}^{2}\right)
\sqrt{\eta^{2}+\mu^{2}}}{r_{+}-r_{-}}+1,\right. 
\left.  \frac{2ir_{+}^{2}\sqrt{\eta^{2}+\mu^{2}}}{r_{+}-r_{-}}%
+1,\frac{2ir_{-}^{2}\sqrt{\eta^{2}+\mu^{2}}}{r_{+}-r_{-}}+1,\sigma
;\frac{2r-\left(  r_{+}+r_{-}\right)  +1}{2\left(  r_{+}-r_{-}\right)
}\right)  ,
\end{align}
\begin{paracol}{2}
\switchcolumn
\noindent where
\end{paracol}
\begin{align}
\qquad\qquad\qquad\sigma = & l\left(  l+1\right)  -2\left(  r_{+}^{2}+r_{+}r_{-}+r_{-}%
^{2}\right)  \eta^{2}-\frac{1}{2}\left(  3r_{+}^{2}+4r_{+}r_{-}+3r_{-}%
^{2}\right)  \mu^{2}\nonumber\\
&  -\eta\left(  r_{+}-r_{-}\right)  \left[  \frac{\left(  r_{+}+r_{-}\right)
\left(  2\eta^{2}+\mu^{2}\right)  }{\eta}-\frac{2r_{+}^{2}\sqrt{\eta^{2}%
+\mu^{2}}}{r_{+}-r_{-}}+i\right] +\frac{\left(  r_{+}^{2}+r_{-}^{2}\right)  \sqrt{\eta^{2}+\mu^{2}}}%
{r_{+}-r_{-}}\left[  \frac{\left(  r_{+}^{2}+r_{-}^{2}\right)  \sqrt{\eta
^{2}+\mu^{2}}}{r_{+}-r_{-}}-i\right]  .
\end{align}
\begin{paracol}{2}
\switchcolumn
\noindent This is the exact scattering wave function. Note that in scattering the
eigenvalue spectrum is a continuous spectrum.

\section{Scattering phase shift: Integral equation \label{Integraleq}}

The above scattering solution is an exact solution. In this section we present
an explicit approximate expression of the scattering phase. We construct an
integral equation by the Green function for the scattering wave function and
solve the scattering phase shift from the integral equation.

\subsection{Effective potential}

The scalar field equation (\ref{radialeq}) is more complicated. In this
section, with the help of the tortoise coordinate, by introducing an effective
potential, we convert the radial equation (\ref{radialeq}) into a
one-dimensional stationary Schr\"{o}dinger equation.

By defining an effective potential%
\begin{equation}
V_{l}^{\text{eff}}=\displaystyle\left(  1-\frac{r_{+}}{r}\right)  \left(
1-\frac{r_{-}}{r}\right)  \left[  \frac{l\left(  l+1\right)  }{r^{2}}+\left(
\frac{r_{+}+r_{-}}{r^{3}}-\frac{2r_{+}r_{-}}{r^{4}}\right)  \right]  +\mu
^{2}\left(  \frac{r_{+}r_{-}}{r^{2}}-\frac{r_{+}+r_{-}}{r}\right)  ,
\end{equation}
we rewrite Eq. (\ref{radialeq1}) as
\begin{equation}
\frac{d^{2}u_{l}\left(  r\right)  }{dr_{\ast}^{2}}+\left(  \eta^{2}%
-V_{l}^{\text{eff}}\right)  u_{l}\left(  r\right)  =0. \label{radialeq2}%
\end{equation}
It can be seen that under the tortoise coordinate $r_{\ast}$ the radial
equation (\ref{radialeq1}) is a one-dimensional stationary Schr\"{o}dinger
equation with the potential $V_{l}^{\text{eff}}$.

Note that for scattering, we only consider the solution at $r>r_{+}$.

\subsection{Green function and integral equation}

In the following we first construct the Green function from the solution of
the radial equation and then construct a scattering integral equation.

First, we construct the Green function from two linearly independent solutions
of the radial equation. Before solving the radial equation (\ref{radialeq2}),
we first solve a solution with $V_{l}^{\text{eff}}=0$:
\begin{equation}
\frac{d^{2}u_{l}^{\left(  0\right)  }\left(  r\right)  }{dr_{\ast}^{2}}%
+\eta^{2}u_{l}^{\left(  0\right)  }\left(  r\right)  =0. \label{radialeq3}%
\end{equation}
It can be checked directly that Eq. (\ref{radialeq3}) has two linearly
independent solutions:%
\begin{align}
y^{\left(  1\right)  }\left(  r\right)   &  =\sin\left(  \eta r_{\ast}\right)
,\label{Lsolution1}\\
y^{\left(  2\right)  }\left(  r\right)   &  =\cos\left(  \eta r_{\ast}\right)
, \label{Lsolution2}%
\end{align}
where the solution is outside the horizon, i.e., $r_{+}<r<\infty$.

By these two linearly independent "free" solutions, we can construct the Green
function \cite{arfken2013mathematical}:%
\begin{align}
G\left(  r,r^{\prime}\right)   &  =C_{1}\left(  r^{\prime}\right)  y^{\left(
1\right)  }\left(  r\right)  +C_{2}\left(  r^{\prime}\right)  y^{\left(
2\right)  }\left(  r\right)  ,\text{ \ }r>r^{\prime},\label{GREEN1}\\
G\left(  r,r^{\prime}\right)   &  =0,\text{
\ \ \ \ \ \ \ \ \ \ \ \ \ \ \ \ \ \ \ \ \ \ \ \ \ }r<r^{\prime}.
\label{GREEN2}%
\end{align}

Now we determine the continuity condition of the Green function.

The Green function for Eq. (\ref{radialeq2}) is defined by
\begin{equation}
\frac{d^{2}G\left(  r,r^{\prime}\right)  }{dr_{\ast}^{2}}+\eta^{2}G\left(
r,r^{\prime}\right)  =\delta\left(  r-r^{\prime}\right)  . \label{Greeneq}%
\end{equation}
Integrating both sides of Eq. (\ref{Greeneq}),%
\begin{equation}
\int_{r^{\prime}-\varepsilon}^{r^{\prime}+\varepsilon}\frac{d^{2}G\left(
r,r^{\prime}\right)  }{dr_{\ast}^{2}}dr+\int_{r^{\prime}-\varepsilon
}^{r^{\prime}+\varepsilon}\eta^{2}G\left(  r,r^{\prime}\right)  dr=\int%
_{r^{\prime}-\varepsilon}^{r^{\prime}+\varepsilon}\delta\left(  r-r^{\prime
}\right)  dr,
\end{equation}
we have%
\begin{equation}
\int_{r^{\prime}-\varepsilon}^{r^{\prime}+\varepsilon}\frac{d}{dr}\left[
\left(  \frac{dr}{dr_{\ast}}\right)  ^{2}\frac{dG\left(  r,r^{\prime}\right)
}{dr}\right]  dr+\int_{r^{\prime}-\varepsilon}^{r^{\prime}+\varepsilon}%
\eta^{2}G\left(  r,r^{\prime}\right)  dr=\int_{r^{\prime}-\varepsilon
}^{r^{\prime}+\varepsilon}\delta\left(  r-r^{\prime}\right)  dr.
\end{equation}
Performing the integral gives%
\begin{equation}
\left.  \left(  \frac{dr}{dr_{\ast}}\right)  ^{2}\frac{dG\left(  r,r^{\prime
}\right)  }{dr_{\ast}}\right\vert _{r^{\prime}-\varepsilon}^{r^{\prime
}+\varepsilon}+\eta^{2}\int_{r^{\prime}-\varepsilon}^{r^{\prime}+\varepsilon
}G\left(  r,r^{\prime}\right)  dr=1. \label{jumpeq}%
\end{equation}
In order to satisfy Eq. (\ref{jumpeq}), we require that $G\left(  r,r^{\prime
}\right)  $ is continuous at $r^{\prime}$ so that the second term vanishes
when $\varepsilon\rightarrow0$, i.e.,%
\begin{equation}
\lim_{\varepsilon\rightarrow0^{+}}\left.  G\left(  r,r^{\prime}\right)
\right\vert _{r=r^{\prime}+\varepsilon}=\lim_{\varepsilon\rightarrow0^{+}%
}\left.  G\left(  r,r^{\prime}\right)  \right\vert _{r=r^{\prime}-\varepsilon
},
\end{equation}
and the first term have a jump satisfying%
\begin{equation}
\lim_{\varepsilon\rightarrow0^{+}}\left.  \left(  \frac{dr}{dr_{\ast}}\right)
^{2}\frac{dG\left(  r,r^{\prime}\right)  }{dr_{\ast}}\right\vert
_{r=r^{\prime}+\varepsilon}-\lim_{\varepsilon\rightarrow0^{+}}\left.  \left(
\frac{dr}{dr_{\ast}}\right)  ^{2}\frac{dG\left(  r,r^{\prime}\right)
}{dr_{\ast}}\right\vert _{r=r^{\prime}-\varepsilon}=1,
\end{equation}
i.e.,%
\begin{equation}
\lim_{\varepsilon\rightarrow0^{+}}\left.  \frac{dG\left(  r,r^{\prime}\right)
}{dr_{\ast}}\right\vert _{r=r^{\prime}+\varepsilon}-\lim_{\varepsilon
\rightarrow0^{+}}\left.  \frac{dG\left(  r,r^{\prime}\right)  }{dr_{\ast}%
}\right\vert _{r=r^{\prime}-\varepsilon}=\frac{1}{\left(  \frac{dr^{\prime}%
}{dr_{\ast}^{\prime}}\right)  ^{2}}.
\end{equation}

Then we arrive at a continuity condition of the Green function:
\begin{align}
\lim_{\epsilon\rightarrow0^{+}}\left.  G\left(  r,r^{\prime}\right)
\right\vert _{r=r^{\prime}+\epsilon}  &  =\lim_{\epsilon\rightarrow0^{+}%
}\left.  G\left(  r,r^{\prime}\right)  \right\vert _{r=r^{\prime}-\epsilon
},\label{GB1}\\
\lim_{\epsilon\rightarrow0^{+}}\left[  \left.  \frac{\partial}{\partial
r}G\left(  r,r^{\prime}\right)  \right\vert _{r=r^{\prime}+\epsilon}-\left.
\frac{\partial}{\partial r}G\left(  r,r^{\prime}\right)  \right\vert
_{r=r^{\prime}-\epsilon}\right]   &  =\frac{1}{\displaystyle\left(
\frac{dr^{\prime}}{dr_{\ast}^{\prime}}\right)  ^{2}}. \label{GB2}%
\end{align}
This gives%
\begin{align}
C_{1}\left(  r^{\prime}\right)  y^{\left(  1\right)  }\left(  r^{\prime
}\right)  +C_{2}\left(  r^{\prime}\right)  y^{\left(  2\right)  }\left(
r^{\prime}\right)   &  =0,\label{green1}\\
C_{1}\left(  r^{\prime}\right)  \left.  \frac{\partial}{\partial r}y^{\left(
1\right)  }\left(  r\right)  \right\vert _{r=r^{\prime}}+C_{2}\left(
r^{\prime}\right)  \left.  \frac{\partial}{\partial r}y^{\left(  2\right)
}\left(  r\right)  \right\vert _{r=r^{\prime}}  &  =\left(  \frac{dr_{\ast
}^{\prime}}{dr^{\prime}}\right)  ^{2}. \label{green2}%
\end{align}
Solving Eqs. (\ref{green1}) and (\ref{green2}) gives the coefficients%
\begin{align}
C_{1}\left(  r^{\prime}\right)   &  =\frac{\cos\left(  \eta r_{\ast}^{\prime
}\right)  }{\eta}\frac{dr_{\ast}^{\prime}}{dr^{\prime}},\\
C_{2}\left(  r^{\prime}\right)   &  =-\frac{\sin\left(  \eta r_{\ast}^{\prime
}\right)  }{\eta}\frac{dr_{\ast}^{\prime}}{dr^{\prime}}.
\end{align}
Then we obtain the Green function
\begin{align}
G\left(  r,r^{\prime}\right)   &  =\frac{\cos\left(  \eta r_{\ast}^{\prime
}\right)  }{\eta}\frac{dr_{\ast}^{\prime}}{dr^{\prime}}\sin\left(  \eta
r_{\ast}\right)  -\frac{\sin\left(  \eta r_{\ast}^{\prime}\right)  }{\eta
}\frac{dr_{\ast}^{\prime}}{dr^{\prime}}\cos\left(  \eta r_{\ast}\right)
,\text{ \ }r>r^{\prime},\label{G1}\\
G\left(  r,r^{\prime}\right)   &  =0,\text{
\ \ \ \ \ \ \ \ \ \ \ \ \ \ \ \ \ \ \ \ \ \ \ \ \ \ \ \ \ \ \ \ \ \ \ \ \ \ \ \ \ \ \ \ }%
r<r^{\prime}. \label{G2}%
\end{align}

By the Green function, we can construct an integral equation for $u_{l}\left(
r\right)  $:
\end{paracol}
\begin{align}
u_{l}\left(  r\right)   &  =Ay^{\left(  1\right)  }\left(  r\right)
+By^{\left(  2\right)  }\left(  r\right)  +\int_{r_{+}}^{r}G\left(
r,r^{\prime}\right)  V_{l}^{\text{eff}}u_{l}\left(  r^{\prime}\right)
dr^{\prime}\nonumber\\
&  =A\sin\left(  \eta r_{\ast}\right)  +B\cos\left(  \eta r_{\ast}\right)
+\frac{1}{\eta}\sin\left(  \eta r_{\ast}\right)  \int_{r_{+}}^{r}\cos\left(
\eta r_{\ast}^{\prime}\right)  \frac{dr_{\ast}^{\prime}}{dr^{\prime}}%
V_{l}^{\text{eff}}u_{l}\left(  r^{\prime}\right)  dr^{\prime} -\frac{1}{\eta}\cos\left(  \eta r_{\ast}\right)  \int_{r_{+}}^{r}%
\sin\left(  \eta r_{\ast}^{\prime}\right)  \frac{dr_{\ast}^{\prime}%
}{dr^{\prime}}V_{l}^{\text{eff}}u_{l}\left(  r^{\prime}\right)  dr^{\prime},
\label{exu}%
\end{align}
\begin{paracol}{2}
\switchcolumn
\noindent or,%
\begin{equation}
u_{l}\left(  r\right)  =A\sin\left(  \eta r_{\ast}\right)  +B\cos\left(  \eta
r_{\ast}\right)  +\frac{1}{\eta}\int_{r_{+}}^{r}\frac{dr_{\ast}^{\prime}%
}{dr^{\prime}}\sin\left(  \eta\left(  r_{\ast}-r_{\ast}^{\prime}\right)
\right)  V_{l}^{\text{eff}}u_{l}\left(  r^{\prime}\right)  dr^{\prime}.
\label{exu1}%
\end{equation}
This is an integral equation of $u_{l}\left(  r\right)  $.

\subsection{Scattering boundary condition}

In this section, we determine the scattering boundary condition by virtue of
the asymptotic behavior of the confluent Heun function.

For scattering, we only concern the large-distance asymptotic behavior of the
radial equation. In the following, we consider the asymptotic equation of the
radial equation.

The confluent Heun equation (\ref{eqyz}), with $p=ic$ and $\beta=i\gamma$, has
an asymptotic solution \cite{ronveaux1995heun}
\begin{equation}
y\left(  z\right)  \overset{z\rightarrow\infty}{\sim}\frac{1}{cz}\sin\left(
cz+\gamma\ln z+O\left(  \frac{1}{z}\right)  \right)  .
\end{equation}
This gives an asymptotic solution of the radial equation (\ref{radialeq4}):%
\end{paracol}
\begin{align}
 \qquad\qquad\qquad\qquad R_{l}\left(  r\right)  \overset{r\rightarrow\infty}{\sim}\frac
{1}{\displaystyle\eta\left[  \left(  r_{+}-r\right)  -\frac{r_{+}-r_{-}}%
{2}\right]  }
 \sin\left(  \eta\left[  \left(  r_{+}-r\right)  -\frac{r_{+}-r_{-}%
}{2}\right]  -\frac{2\eta^{2}+\mu^{2}}{2\eta}\left(  r_{+}+r_{-}\right)
\ln\left(  1-2\frac{r_{+}-r}{r_{+}-r_{-}}\right)  \right)  . \label{asR}%
\end{align}
\begin{paracol}{2}
\switchcolumn
\noindent For high energy scattering, $\mu/\eta\ll1$, the asymptotics of the radial
function can be written as
\end{paracol}
\begin{align}
R_{l}\left(  r\right)  \overset{r\rightarrow\infty}{\sim}\frac
{1}{\displaystyle\eta\left[  \left(  r-r_{+}\right)  +\frac{r_{+}-r_{-}}%
{2}\right]  }\sin\left(  \eta\left[  r+\left(  r_{+}+r_{-}\right)  \ln\left(
\frac{r}{r_{+}+r_{-}}-1\right)  +\left(  r_{+}+r_{-}\right)  \ln2\right.
\right. 
\left.  \left.  -\left(  r_{+}+r_{-}\right)  \ln\frac{r_{+}-r_{-}}%
{r_{+}+r_{-}}-\frac{r_{+}+r_{-}}{2}\right]  +\delta_{l}-\frac{l\pi}{2}\right)
. \label{asR1}%
\end{align}
\begin{paracol}{2}
\switchcolumn
\noindent
Introducing $R_{l}\left(  r\right)  =u_{l}\left(  r\right)  /r$, we arrive at%
\begin{equation}
u_{l}\left(  r\right)  \sim\sin\left(  \eta\left[  r+\left(  r_{+}%
+r_{-}\right)  \ln\left(  \frac{r}{r_{+}+r_{-}}-1\right)  \right]  +\delta
_{l}-\left(  r_{+}+r_{-}\right)  \eta\ln\frac{r_{+}-r_{-}}{r_{+}+r_{-}}%
+\Delta\left(  \eta\right)  \right)  , \label{asu1}%
\end{equation}
where $\delta_{l}$ is the scattering phase shift and%
\begin{align}
\Delta\left(  \eta\right)   &  =-\frac{l\pi}{2}-\frac{r_{+}+r_{-}}{2}%
\eta+\left(  r_{+}+r_{-}\right)  \eta\ln2\nonumber\\
&  =-\frac{l\pi}{2}-M\eta+2M\eta\ln2.
\end{align}

Now we obtain the asymptotic solution. This requires that the asymptotics of
the solution of the radial equation (\ref{radialeq}) should take the form of
the asymptotics (\ref{asu1}); that is, Eq. (\ref{asu1}) is just the scattering
boundary condition.

\subsection{Scattering phase shift and scattering cross section
\label{phaseshift}}

Now we calculate the scattering phase shift.

In order to compare to the asymptotic solution (\ref{asu1}), we seek for an
asymptotic equation of the integral equation (\ref{exu1}). Rewrite the
integral equation (\ref{exu1}) as
\end{paracol}
\begin{align}
\qquad\qquad\qquad u_{l}\left(  r\right)    =\left(  A+\frac{1}{\eta}\int_{r_{+}}^{r}%
\cos\left(  \eta r_{\ast}^{\prime}\right)  \frac{dr_{\ast}^{\prime}%
}{dr^{\prime}}V_{l}^{\text{eff}}u_{l}\left(  r^{\prime}\right)  dr^{\prime
}\right)  \sin\left(  \eta r_{\ast}\right) +\left(  B-\frac{1}{\eta}\int_{r_{+}}^{r}\sin\left(  \eta r_{\ast}^{\prime
}\right)  \frac{dr_{\ast}^{\prime}}{dr^{\prime}}V_{l}^{\text{eff}}u_{l}\left(
r^{\prime}\right)  dr^{\prime}\right)  \cos\left(  \eta r_{\ast}\right)  .
\end{align}
\begin{paracol}{2}
\switchcolumn
\noindent Then taking $r\rightarrow\infty$ gives the asymptotics%
\begin{align}
u_{l}\left(  r\right)  \overset{r\rightarrow\infty}{\sim}\alpha\left(
\eta\right)  \sin\left(  \eta r_{\ast}\right)  +\beta\left(  \eta\right)
\cos\left(  \eta r_{\ast}\right)  =C\sin\left(  \eta r_{\ast}+\phi\right)  , \label{exasu1}%
\end{align}
where%
\begin{align}
\alpha\left(  \eta\right)   &  =A+\frac{1}{\eta}\int_{r_{+}}^{\infty}%
\cos\left(  \eta r_{\ast}\right)  \frac{dr_{\ast}}{dr}V_{l}^{\text{eff}}%
u_{l}\left(  r\right)  dr,\\
\beta\left(  \eta\right)   &  =B-\frac{1}{\eta}\int_{r_{+}}^{\infty}%
\sin\left(  \eta r_{\ast}\right)  \frac{dr_{\ast}}{dr}V_{l}^{\text{eff}}%
u_{l}\left(  r\right)  dr,
\end{align}%
\begin{align}
\tan\phi   =\frac{\beta\left(  \eta\right)  }{\alpha\left(  \eta\right)  },\text{\quad}
\cos\phi   =\frac{\alpha\left(  \eta\right)  }{\sqrt{\alpha^{2}\left(
\eta\right)  +\beta^{2}\left(  \eta\right)  }},\text{\quad }
\sin\phi   =\frac{\beta\left(  \eta\right)  }{\sqrt{\alpha^{2}\left(
\eta\right)  +\beta^{2}\left(  \eta\right)  }},
\end{align}
and%
\begin{equation}
C=\sqrt{\alpha^{2}\left(  \eta\right)  +\beta^{2}\left(  \eta\right)  }.
\end{equation}
By the tortoise coordinate (\ref{tortoise1}), Eq. (\ref{exasu1}) can be
written as%
\begin{equation}
u_{l}\left(  r\right)  \overset{r\rightarrow\infty}{\sim}C\sin\left(
\eta\left[  r+\frac{r_{+}^{2}}{r_{+}-r_{-}}\ln\left(  \frac{r}{r_{+}%
}-1\right)  -\frac{r_{-}^{2}}{r_{+}-r_{-}}\ln\left(  \frac{r}{r_{-}}-1\right)
\right]  +\phi\right)  . \label{exasu2}%
\end{equation}

The scattering phase shift can be obtained by comparing two asymptotic
solutions, Eqs. (\ref{asu1}) and (\ref{exasu2}). Direct comparison gives%
\begin{equation}
\phi=\delta_{l}-\left(  r_{+}+r_{-}\right)  \eta\ln\frac{r_{+}-r_{-}}%
{r_{+}+r_{-}}+\Delta\left(  \eta\right)  .
\end{equation}
The scattering phase shift then reads%
\end{paracol}
\begin{align}
\qquad\qquad\qquad\qquad\qquad\qquad\delta_{l}  &  =\arctan\frac{\beta\left(  \eta\right)  }{\alpha\left(
\eta\right)  }+\left(  r_{+}+r_{-}\right)  \eta\ln\frac{r_{+}-r_{-}}%
{r_{+}+r_{-}}-\Delta\left(  \eta\right) \nonumber\\
&  =\arctan\frac{B-\displaystyle\frac{1}{\eta}\int_{r_{+}}^{\infty}\sin\left(
\eta r_{\ast}\right)  \frac{dr_{\ast}}{dr}V_{l}^{\text{eff}}u_{l}\left(
r\right)  dr}{A+\displaystyle\frac{1}{\eta}\int_{r_{+}}^{\infty}\cos\left(
\eta r_{\ast}\right)  \frac{dr_{\ast}}{dr}V_{l}^{\text{eff}}u_{l}\left(
r\right)  dr}+\left(  r_{+}+r_{-}\right)  \eta\ln\frac{r_{+}-r_{-}}{r_{+}+r_{-}}%
-\Delta\left(  \eta\right)  , \label{ps}%
\end{align}
\begin{paracol}{2}
\switchcolumn
\noindent where $\displaystyle\phi=\arctan\frac{\beta\left(  \eta\right)  }%
{\alpha\left(  \eta\right)  }$ is used.

Now we determine the constants $A$ and $B$.

For $V_{l}^{\text{eff}}=0$ with $M=0$ and$\ Q=0$, Eq. (\ref{exu1}) becomes%
\begin{equation}
u_{l}\left(  r\right)  =A\sin\left(  \eta r\right)  +B\cos\left(  \eta
r\right)  .
\end{equation}
The boundary condition requires that at the horizon $r=r_{+}$ with
$r_{+}=M+\sqrt{M^{2}-Q^{2}}=0$, we must have $u_{l}\left(  0\right)  =0$ so
that the radial wave function is finite. Consequently, $B=0$ and $A$ is an
arbitrary constant. Taking $A=1$ gives the scattering phase shift
\end{paracol}
\newpage
\begin{align}
\qquad\qquad\qquad\qquad\qquad\delta_{l}  &  =-\arctan\frac{\displaystyle\frac{1}{\eta}\int_{r_{+}}^{\infty
}\sin\left(  \eta r_{\ast}\right)  \frac{dr_{\ast}}{dr}V_{l}^{\text{eff}}%
u_{l}\left(  r\right)  dr}{1+\displaystyle\frac{1}{\eta}\int_{r_{+}}^{\infty
}\cos\left(  \eta r_{\ast}\right)  \frac{dr_{\ast}}{dr}V_{l}^{\text{eff}}%
u_{l}\left(  r\right)  dr}+\left(  r_{+}+r_{-}\right)  \eta\ln\frac{r_{+}-r_{-}}{r_{+}+r_{-}}%
+\frac{l\pi}{2}+\frac{r_{+}+r_{-}}{2}\eta-\left(  r_{+}+r_{-}\right)  \eta
\ln2\nonumber\\
&  =-\arctan\frac{\displaystyle\frac{1}{\eta}\int_{r_{+}}^{\infty}\sin\left(
\eta r_{\ast}\right)  \frac{dr_{\ast}}{dr}V_{l}^{\text{eff}}u_{l}\left(
r\right)  dr}{1+\displaystyle\frac{1}{\eta}\int_{r_{+}}^{\infty}\cos\left(
\eta r_{\ast}\right)  \frac{dr_{\ast}}{dr}V_{l}^{\text{eff}}u_{l}\left(
r\right)  dr}∫+\eta\left(  r_{+}+r_{-}\right)  \ln\frac{r_{+}-r_{-}}{r_{+}+r_{-}}%
+\frac{l\pi}{2}+M\eta-2M\eta\ln2. \label{psp}%
\end{align}

\begin{paracol}{2}
\switchcolumn
The zeroth-order phase shift is then
\begin{align}
\delta_{l}^{\left(  0\right)  }  &  =\frac{l\pi}{2}+\frac{r_{+}+r_{-}}{2}%
\eta-\left(  r_{+}+r_{-}\right)  \eta\ln2\nonumber\\
&  =\frac{l\pi}{2}+M\eta-2M\eta\ln2+2M\eta\ln\frac{\sqrt{M^{2}-Q^{2}}}{M}.
\label{ps0}%
\end{align}
The first-order phase shift, by substituting the zeroth-order wave function
\[
u_{l}^{\left(  0\right)  }\left(  r\right)  =\sin\left(  \eta r_{\ast}\right)
\]
into Eq. (\ref{psp}), reads
\begin{align}
\delta_{l}^{\left(  1\right)  }  &  =-\arctan\frac{\displaystyle\frac{1}{\eta
}\int_{r_{+}}^{\infty}\sin^{2}\left(  \eta r_{\ast}\right)  \frac{dr_{\ast}%
}{dr}V_{l}^{\text{eff}}dr}{1+\displaystyle\frac{1}{\eta}\int_{r_{+}}^{\infty
}\sin\left(  2\eta r_{\ast}\right)  \frac{dr_{\ast}}{dr}V_{l}^{\text{eff}}%
dr}+\left(  r_{+}+r_{-}\right)  \eta\ln\frac{r_{+}-r_{-}}{r_{+}+r_{-}%
}\nonumber\\
&  =-\arctan\frac{\displaystyle\frac{1}{\eta}\int_{r_{+}}^{\infty}\sin
^{2}\left(  \eta r_{\ast}\right)  \frac{dr_{\ast}}{dr}V_{l}^{\text{eff}}%
dr}{1+\displaystyle\frac{1}{\eta}\int_{r_{+}}^{\infty}\sin\left(  2\eta
r_{\ast}\right)  \frac{dr_{\ast}}{dr}V_{l}^{\text{eff}}dr}+\left(  r_{+}%
+r_{-}\right)  \eta\ln\frac{r_{+}-r_{-}}{r_{+}+r_{-}}. \label{ps1}%
\end{align}

The scattering amplitude then reads \cite{pike2008scatteringPage}%
\begin{equation}
f\left(  \theta\right)  =\frac{1}{2i\omega}\sum_{l=0}^{\infty}\left(
2l+1\right)  \left(  e^{2i\delta_{l}}-1\right)  P_{l}\left(  \cos
\theta\right)  , \label{fTheta}%
\end{equation}
and the differential scattering cross section is
\[
\sigma\left(  \theta\right)  =\left\vert f\left(  \theta\right)  \right\vert
^{2}.
\]

In Fig. (\ref{sigma}), we compare three differential scattering cross sections
up to the second-order scattering phase shift: the Schwarzschild case
$Q/M\simeq0$, the typical Reissner-Nordstr\"{o}m case $Q/M=1/2$, and the
extremal Reissner-Nordstr\"{o}m case $Q/M\simeq1$. Here the phase shift
$\delta_{l}\simeq\delta_{l}^{\left(  0\right)  }+\delta_{l}^{\left(  1\right)
}$ and sums to $l=6$.

\medskip
The scattering amplitude $f\left(  \theta\right)  $ is given by the series
(\ref{fTheta}). The scattering amplitude $f\left(  \theta\right)  $ in the
interval from $0$ to $\pi$ has no oscillations. However, often, the sum in Eq.
(\ref{fTheta}) cannot be performed exactly, and one has to approximately
replace the exact sum by a partial sum consisting of the first several terms.
\ The partial sum has an incorrect oscillation which does not appear in the
exact sum. The incorrect oscillation cannot be eliminated by simply keeping
more terms into account. We suggest an approach to eliminate such an
oscillation in the partial sum approximation in \cite{li2021eliminating}. 

\vspace*{-0.5cm}
\begin{wrapfigure}[18]{l}[0.5cm]{0.6\textwidth}
\includegraphics[width=0.6\textwidth]{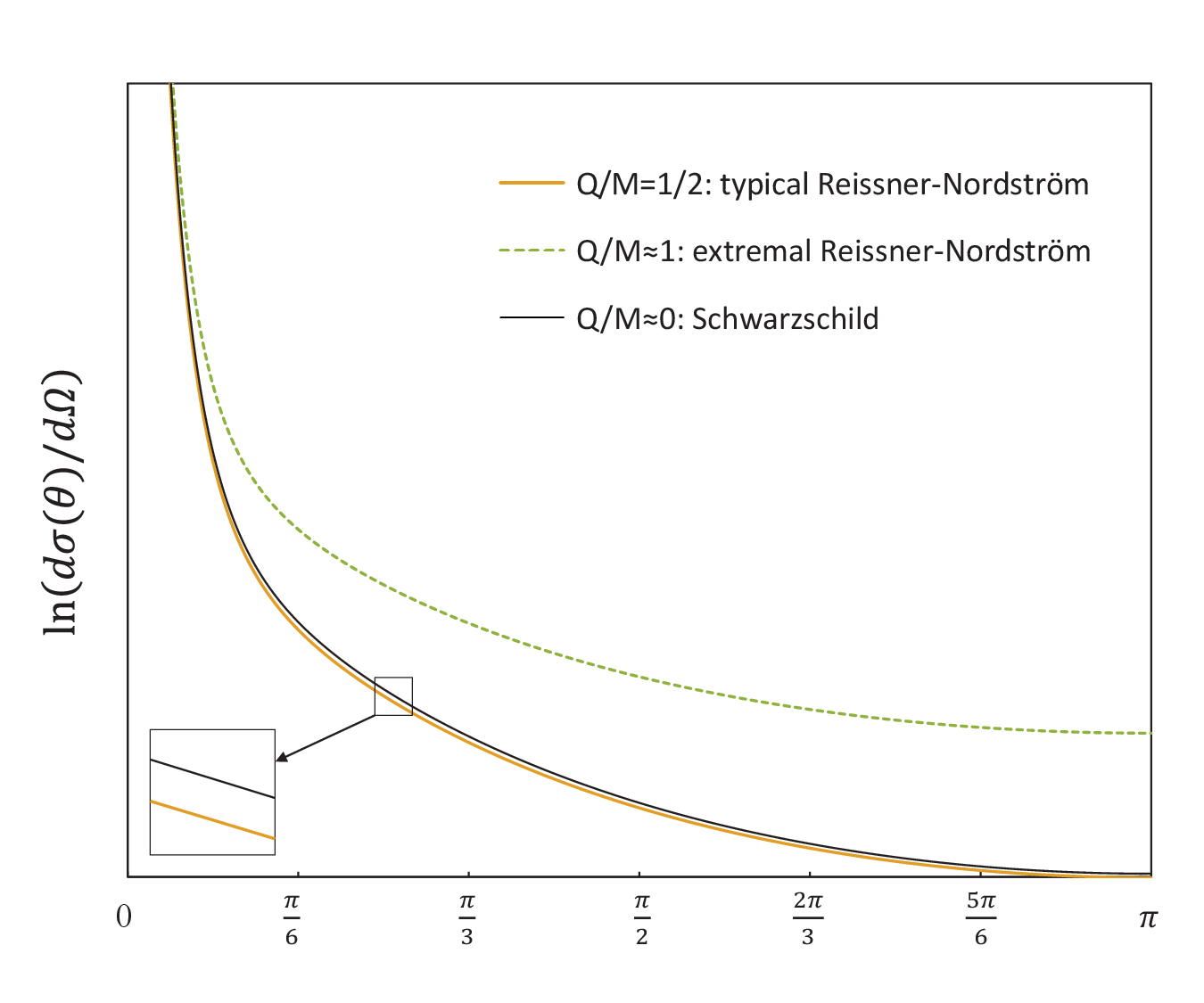}
\captionsetup{labelformat=empty} 
\caption{} 
\label{sigma} 
\end{wrapfigure}
\
\qquad\qquad\\
\qquad\qquad\\
\qquad\qquad\\
\qquad\qquad\\
\qquad\qquad\\
\qquad\qquad\\
\qquad\qquad\\
\qquad\qquad\\
\qquad\qquad\\
\vspace*{0.5cm}
\qquad\qquad\\
\qquad\qquad\\
\qquad\qquad\\
\qquad\qquad\\
\qquad\qquad\\
\qquad\qquad\\
\qquad\qquad\\

\bigskip
\switchcolumn
\newpage
\vspace*{1.5cm}
{\color{Gray} 
\noindent\textbf{Figure \ref{sigma}.} Differential scattering cross sections of the typical Reissner-Nordstr\"{o}m
case $Q/M=1/2$, the extremal Reissner-Nordstr\"{o}m case $Q/M\simeq1$, and the
Schwarzschild case $Q/M\simeq0$. 
}
\switchcolumn

\vspace*{1.0cm}

\section{Conclusion \label{Conclusions}}

In this paper, we solve the massive scalar field in the Reissner-Nordstr\"{o}m
spacetime. The solutions of bound states and scattering states are presented.
The bound-state wave function, the bound-state eigenvalue and the scattering
wave function are calculated by directly solving the radial equation. In order
to obtain an explicit expression of the scattering phase shift, we use the
integral approach.

The scattering boundary condition of long range potentials is difficult to
determine, since different long range potentials have different scattering
boundary conditions. Scattering on the Reissner-Nordstr\"{o}m spacetime is
essentially a long range potential. In this paper, we determine the scattering
boundary condition based on the asymptotic behavior of the confluent Heun function.

Moreover, it is worthy to note here that the result given in the present paper
recovers the result of the Schwarzschild spacetime when the charge $Q=0$.

In the calculation of scattering cross sections, we encounter an incorrect
oscillation in the partial sum approximation. We suggest an approach for
eliminating such oscillations in the appendix.

The scattering phase shift also plays an important role in quantum field
theory through the scattering spectral method \cite{graham2009spectral}. The
heat kernel method is another important method in quantum field theory
\cite{barvinsky1987beyond,barvinsky1990covariant,barvinsky1990covariant3,dai2009number,dai2010approach,mukhanov2007introduction}%
. In virtue of the relation between the scattering spectral method and the
heat kernel method \cite{pang2012relation,li2015heat}, the result of the
scattering phase shift can also be applied to the heat kernel theory in
quantum field theory.

\bigskip
\bigskip
\bigskip



\acknowledgments

We are very indebted to Dr G. Zeitrauman for his encouragement. This work is supported in part by Special Funds for theoretical physics Research Program of the NSFC under Grant No.
11947124, and NSFC under Grant Nos. 11575125 and 11675119.

\nolinenumbers

\reftitle{References}




\providecommand{\href}[2]{#2}\begingroup\raggedright\endgroup

\end{paracol}
%


\end{document}